\documentclass[prd,tightenlines,showpacs,letterpaper, preprintnumbers,nofootinbib,floatfix]{revtex4}

\usepackage{amsmath,amssymb,amsfonts,graphicx,pifont,dcolumn}
\usepackage{epstopdf}
\usepackage{color}
\usepackage{cancel}
\usepackage{ulem}
\usepackage{xfrac}
\RequirePackage{slashed}

\begin{document}

\title{Energy Loss of Monopolium in a Medium}

\author{Huner Fanchiotti}  \author{Carlos A.~Garc\'\i a Canal}
\affiliation{ IFLP(CONICET) and Department of Physics, University of La Plata,  C.C. 67 1900, La Plata, Argentina}

\author{ Vicente Vento }

\affiliation{Departamento de F\'{\i}sica Te\'orica and IFIC, Universidad de Valencia - CSIC, E-46100 Burjassot (Valencia), Spain.}

\date{\today}

\begin{abstract}
We study the energy loss of excited monopolium in an atomic medium. We perform a classical calculation in line with a similar calculation performed for charged particles which leads in the non relativistic limit to the Bethe-Bloch formula except for the density dependence of the medium, which we do not consider in this paper. Our result shows that for maximally deformed  Rydberg states the ionization of monopolium in a light atomic medium is similar to that of light ions.

\end{abstract}


\maketitle

\section{Introduction}

Inspired by the old idea of Dirac and Zeldovich~\cite{Dirac:1931qs,Zeldovich:1978wj} that monopoles are not seen free because they are confined by their strong magnetic forces, we have studied a monopole-antimonopole bound state that we have called monopolium~\cite{Vento:2007vy}. This state is the strongly coupled dual analog of positronium and decays into photons \cite{Epele:2012jn}. The discovery of monopole and dipole solutions in Kaluza Klein theories \cite{Kaluza:1921tu,Klein:1926tv} made these theories very exciting
from a theoretical point of view \cite{Gross:1983hb}. In particular the dipole solution, which is classically stable and therefore very long lived, forms a very
interesting state which we have also called monopolium in analogy with its decaying cousin in  gauge theories. The Kaluza Klein monopolium is extremely massive with a mass of the order of the Planck mass and therefore impossible to produce in laboratories. However, there might be clouds of them in the cosmos which might enter our detectors~\cite{Vento:2020vsq}.

Much experimental research has been carried out into the search for monopoles \cite{Cabrera:1982gz,Milton:2006cp,MoEDAL:2014ttp,Acharya:2014nyr,Patrizii:2015uea}. However, our interest  lies in the detection of monopolium, which is chargeless but  can manifest a magnetic moment in the presence of magnetic fields (dual Stark effect) in excited deformed states \cite{Vento:2020vsq,Vento:2019auh,Fanchiotti:2022xvx}.  In this work we want to analyze the capacity of monopolium to produce ionization when traversing a medium by calculating its energy loss when colliding with atomic electrons. The idea behind this investigation is to find out if TPCs \cite{Nygren:2018sjx} could in principle be used for detecting monopolium. 

We start by reviewing in section \ref{charge}  a  classical calculation of the energy loss by a heavy charged particle, a well known problem, that will serve to present the formalism that will be later used in the case of monopolium. This classical result provides some of the dependences that describe the stopping power of a heavy charged particle in a medium. Thereafter, in section \ref{energyloss} we follow the same procedure to calculate the stopping power for a heavy particle with no charge but with magnetic moment. We next compare the result for the charged particle and neutral magnetic by establishing a ratio of energy losses. In section \ref{monopolium} we apply the described analysis to monopolium. We end up by some concluding remarks.

\section{Energy Loss by a heavy charged particle traversing a medium}
\label{charge}

A heavy charged particle traversing a medium interacts  electromagnetically mainly with the atomic electrons in the medium. In the process it transfers energy to them ionizing the atoms of the medium. The quantum mechanical calculation is quite complicated and leads to the famous Bethe-Bloch formula \cite{Bethe:1930ku,Bloch:1933zza}. However, since we would like to compare this case with the case of monopolium we will resorts to the much simpler classical calculation which shows the main dependencies of the process \cite{Foudas:2007cu,Ahlen:1976jw,Ahlen:1982mx}.

\begin{figure}[htb]
\begin{center}
\includegraphics[scale= 0.8]{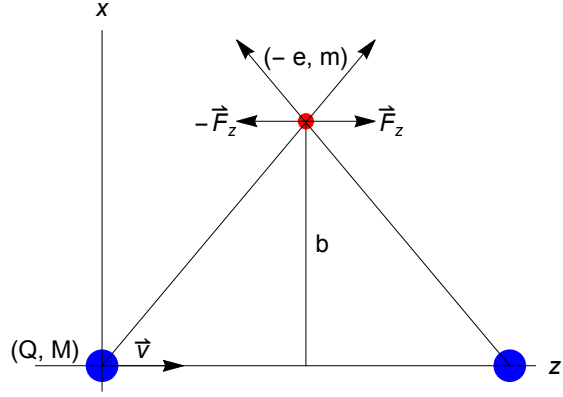} 
\end{center}
\caption{Heavy charged particle traveling with velocity $v$ along the $z$ direction passing at impact parameter $b$  from an atomic electron.}
\label{Ionizationcharge}
\end{figure}

Let us solve the problem of calculating the energy transfer from a heavy charged particle that moves with velocity $v<<c$ towards a stable electron (see Fig. \ref{Ionizationcharge}). Since we want to calculate the energy loss $dE/dz$ the distance traveled $dz$ is small, thus to keep $v$ constant is a good approximation.

The force acting on the electron is 

\begin{equation}
\vec{F} = \frac{Q e \hat{r}}{|\vec{r}(t)|^2} = F_x \hat{x} + F_z \hat{z}.
\end{equation}
Due to the symmetry of the problem $F_z$ cancels out as shown in Fig. \ref{Ionizationcharge} and $F_x$ is

\begin{equation}
F_x= \frac{Q e \sin{\theta}}{|\vec{r}(t)|^2} = \frac{Q e b}{|\vec{r}(t)|^3} = \frac{Q e b}{(b^2 + (v t)^2)^{\sfrac{3}{2}}},
\end{equation}
where $\theta$ is the angle between $\vec{F}$ and the $z$ axis. From $F_x$ we get the velocity

\begin{equation}
v_x = \frac{Q e}{m} \int_{-\infty}^{-\infty} \frac{b}{(b^2 + (v t)^2)^{\sfrac{3}{2}}} dt = \frac{2 Q e}{m v b},
\end{equation}
where $v=|\vec{v}|$.  

The energy gained by the electron is

\begin{equation}
\Delta E = \frac{1}{2} m v_x^2 = \frac{2 Q^2 e^2}{m v^2 b^2}.
\label{Eb}
\end{equation}

\begin{figure}[htb]
\begin{center}
\includegraphics[scale= 0.8]{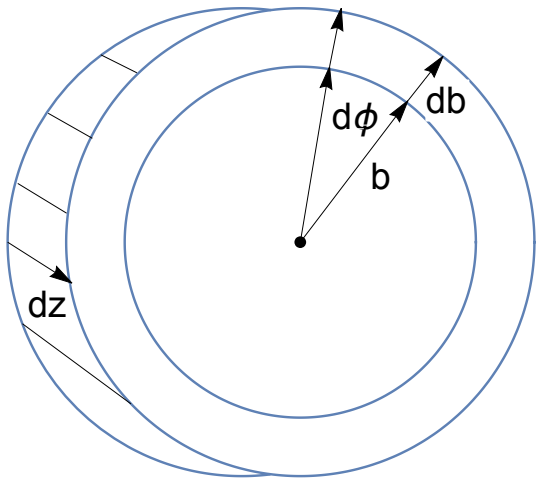} 
\end{center}
\caption{The elementary Integration volume to obtain the interaction with all the electrons}
\label{volume1}
\end{figure}

In order to obtain the contribution of all the electrons interacting with the heavy particle we need to consider the cylindrical volume in Fig. \ref{volume1}.

\begin{equation}
dV = 2 \pi b db dz.
\label{eqvolume}
\end{equation}
Note that although we have calculated the interaction in the plane $x-z$ the interaction has cylindrical symmetry and the active force goes into the radial direction in all planes.
In order to calculate the total energy loss we have to take into account all the electrons per unit volume and multiply by N  the electron number density.

Thus the total energy loss is this given by

\begin{equation}
\frac{dE}{dz} = 2 \pi \frac{Q^2 e^2 N}{m v^2} \log{\frac{b_{max}}{b_{min}}}
\label{energylossclassical}
\end{equation}

Given the relation between the impact parameter and the energy, Eq.(\ref{Eb}), we obtain for the energy loss

\begin{equation}
\frac{dE}{dz} = 2 \pi \frac{Q^2 e^2 N}{m v^2} \log{\frac{E_{max}}{E_{min}}}.
\end{equation}
The maximum energy to leading order in $(m/M)$ is given by \cite{Ahlen:1976jw} 

\begin{equation}
E_{max}= \frac{2 m c^2 \beta^2}{1- \beta^2}
\end{equation}
where $\beta = v/c$, and therefore

\begin{equation}
\frac{dE}{dz} = 2 \pi \frac{Q^2 e^2 N}{m c^2 \beta^2}  \log{\frac{2 m \beta^2}{(1-\beta^2)E_{min}}}.
\label{classical}
\end{equation}
which agrees with the more sophisticated quantum calculation by Ahlen \cite{Ahlen:1976jw} for small velocity $Z/\beta>137$. This result is an approximation to the more 
complete Bethe-Bloch \cite{Bethe:1930ku,Bloch:1933zza} relation which includes the effects of atomic or molecular binding and relativistic corrections, 

\begin{equation}
\frac{dE}{dz} = 2 \pi \frac{Q^2 e^2 N}{m c^2 \beta^2}  \left(\log{\frac{2 m \beta^2}{(1-\beta^2)E_{min}}} - \beta^2 -\delta_e/2\right),
\label{bethebloch}
\end{equation}
where $\delta_e$ characterizes the density effect and $E_{min}$ is the mean ionization potential which can be approximated by $10 Z$ eV~\cite{Foudas:2007cu,Ahlen:1976jw}, where $Z$ is the atomic number of the atom in the medium . In this initial attempt our aim is to look at the behavior of the first term which is dominant for very small velocity and disregard the density dependence. Our comparison will be valid therefore only in the kinematical region where the first term of Eq.(\ref{bethebloch}) is dominant.

Note that we use $4\pi \epsilon_0 =1$, $ \mu_0/4\pi = 1$ and therefore $c=1$. Sometimes though we shall keep explicitly $c$ during the calculation process, to respect the notation of the classical equations, although at the end we will obtain all our equations in natural units.

\section{Energy Loss by  heavy neutral particles with magnetic moment in a medium}
\label{energyloss}

Our interest lies in the possibility of detecting monopolium \cite{Vento:2007vy},  a bound monopole anti-monopole state, by its ionization. Monopolium is neutral and is heavy, but in its excited states it has a magnetic moment in the presence of even weak magnetic fields \cite{Vento:2020vsq,Fanchiotti:2022xvx}. Heavy neutral particles with magnetic moment (this work) or electric dipole moment~\cite{PhysRevA.65.022902} interact electromagnetically mainly with the atomic electrons in the medium and transfer energy to them in a process which results in ionizing the atoms of the medium. Calculating in this case  the energy loss in a medium using quantum theory is quite complicated,  however we have just shown that for small velocities the classical calculation agrees well with the quantum mechanical one. The classical calculation demonstrates the origin of the main features and dependencies of the energy loss. Let us proceed therefore by developing the classical dynamics of the process as shown in Fig. \ref{Ionizationcharge} changing $(Q, M)$ by $(\overrightarrow{\mathcal M},M)$ and $(-e, m)$ by $(-e,\vec{\mu},m)$.

We begin by describing the dynamics of the collision between a moving neutral heavy particle with magnetic moment and an atomic electron.
We assume a neutral heavy particle  with magnetic moment $\overrightarrow{\mathcal M}$ moving with velocity $\vec{v}$ along the $z$ direction. In its own frame of reference it will create a magnetic field 

\begin{align}
 \overrightarrow{B_M} = &\left(\frac{3 \vec{r}\cdot\overrightarrow{\mathcal M}\, \vec{r}}{r^5}-\frac{\overrightarrow{\mathcal M}}{r^3}\right), 
 \label{magneticMfield}
\end{align}

The electron has charge $-e$ and magnetic moment $\vec{\mu}$ therefore in its own frame creates an electric field

\begin{equation}
\overrightarrow{E_e} =-\frac{e}{r^2} \,\hat{r}
\end{equation}
and a magnetic field

\begin{align}
 \overrightarrow {B_e} = &\left(\frac{3 \vec{r}\cdot\vec{\mu}\, \vec{r}}{r^5}-\frac{\vec{\mu}}{r^3}\right). 
\end{align}

In the frame of the electron the heavy particle has both electric and magnetic fields as dictated by the Lorentz transformation from the heavy particle reference system $S$ to the electron reference system $S^\prime$, which is given by

\begin{align}
E^\prime_{M z} & = 0 & E^\prime_{M x}& = -\gamma \beta B_{M y} & E^\prime_{M y}& = \gamma \beta B_{M x},
\label{ecase}
\end{align}
and a magnetic field
\begin{align}
B^\prime_{M z} & = B_{M z} & B^\prime_{M x} & = \gamma B_{M x} & B^\prime_{M y}& = \gamma  B_{M y},
\label{mucase}
\end{align}
where $\gamma = 1/\sqrt{1- \beta^2}$. In the following subsections we calculate the contribution to the energy loss of a heavy moving particle with magnetic moment taking into account the interaction of the electron charge with this electric field and that of the electron magnetic moment with the magnetic field.

We expect slow particles to ionize more like in the charge case thus a non-relativistic limit in the case of a very heavy particle is reasonable. From Eqs.(\ref{ecase}) and (\ref{mucase}) it would seem that for low velocities  the interaction of the magnetic moment of the electron $\vec{\mu}$ with the magnetic field of the heavy particle Eq.(\ref{mucase}), $\overrightarrow{B_{M}}$ is $\beta$ dominant. We will see that this only happens for extremely low velocities due to the dependence on the distance of the interactions.

\subsection{Energy Loss produced by the  Magnetic field interaction}
Let us start by calculating the energy loss of the magnetic field interaction. The interaction in this case is given by

\begin{equation}
H = - \vec{\mu}\cdot \overrightarrow{B_M}
\end{equation}
and following the classical description of the previous section we have a force acting on the electron,

\begin{equation}
\overrightarrow{F_e}= -\vec{\nabla} H =\vec{\nabla} \left( \frac{3 \vec{r}\cdot\overrightarrow{{\mathcal M}} \vec{r}\cdot \vec{\mu}}{r^5} -\frac{\overrightarrow{{\mathcal M}}\cdot\vec{\mu}}{r^3}\right).
\end{equation}

Thus

\begin{equation}
F_{e i} = 3\frac{{\mathcal M}_i \vec{r}\cdot\vec{\mu} + \mu_i \vec{r}\cdot \overrightarrow{{\mathcal M}}}{r^5} -15 \frac{r_i \vec{r}\cdot\overrightarrow{{\mathcal M}} \vec{r}\cdot \vec{\mu}}{r^7} +5\frac{r_i\overrightarrow{{\mathcal M}}\cdot\vec{\mu}}{r^5}.
\end{equation}
where $i=x,y,z$ and $r_x=x, r_y = y$ and $ r_z =z$.

Fig.\ref{Ionizationcharge} shows also now that $F_{ez}$ will vanish and therefore we are left with  $F_{ex}$ and $F_{ey}$, and with the corresponding velocities  $v_{ex}$ and $v_{ey}$. 

Recall that $r =\sqrt{b^2+ (vt)^2}$ and $\vec{r} = vt \hat{k} + b \hat{\rho}= vt \hat{k} + b \cos{\phi} \hat{i} + b \sin{\phi} \hat{j}$, where $\hat{i}, \hat{j},\hat{k}$ form the cartesian unit basis, $\phi$ is the polar angle in the $x,y$ plane and we use $c \ne 1$ until further notice. Let us calculate the velocity in an arbitrary point in the circle of radius $b$ in the $x-y$ plane $(b \cos{\phi}, b\sin{\phi})$, whose components in the plane are given by

\begin{eqnarray}
m v_{e x} = &\int_\infty^\infty F_{e x} dt \nonumber\\
m v_{e y} = &\int_\infty^\infty F_{e y} dt 
\end{eqnarray}

All the integrals can be easily performed and the results is

\begin{eqnarray}
v_{e x} &= &\frac{1}{4 m v b^3} ( (2{\mathcal M}_x \mu_x -  {\mathcal M}_z \mu_z + \frac{5}{3} \overrightarrow{{\mathcal M}} \cdot \vec{\mu})\cos{\phi} +({\mathcal M}_x \mu_y + {\mathcal M}_y \mu_x) \sin{\phi} \nonumber\\
&&\; \; \; \; \; \; \; \; \; \; \; \;- 4({\mathcal M}_y \mu_y \cos{\phi} \sin^2{\phi} + ({\mathcal M}_x \mu_y + {\mathcal M}_y \mu_x ) \cos^2{\phi}\sin{\phi}+{\mathcal M}_x \mu_x \cos^3{\phi}))= \frac{4 {\mathcal M}\mu_b}{m v b^3} \Phi_{\mu x},\\
v_{e y} &= &\frac{1}{4m v b^3} ( (2{\mathcal M}_y \mu_y -  {\mathcal M}_z \mu_z + \frac{5}{3} \overrightarrow{{\mathcal M}} \cdot \vec{\mu})\sin{\phi} +({\mathcal M}_x \mu_y + {\mathcal M}_y \mu_x) \cos{\phi} \nonumber\\
&&\; \; \; \; \; \; \; \; \; \; \; \;- 4({\mathcal M}_y \mu_y  \sin^3{\phi} + ({\mathcal M}_x \mu_y + {\mathcal M}_y \mu_x ) \cos{\phi}\sin^2{\phi}+{\mathcal M}_x \mu_x \cos^2{\phi} \sin{\phi}))=\frac{4 {\mathcal M}\mu_b}{m v b^3} \Phi_{\mu y}. 
\label{velocityM}
\end{eqnarray}
These equations define $\Phi_{\mu x}$ and $ \Phi_{\mu y}$ which enter the deposited energy as 

\begin{equation}
E = \frac{8{\mathcal M}^2 \mu_B^2}{m v^2 b^6} (\Phi_{\mu x}^2 + \Phi_{\mu y}^2)
\label{energyM}
\end{equation}
where $\mu_B$ is Bohr's magneton and $\Phi_{\mu i}^2,  i = x, y$ are the angular dependences determined from $v_{e i}$ by substituting  $\overrightarrow{{\mathcal M}} = {\mathcal M}( \sin{\theta_{{\mathcal M}}}\cos{\phi_{{\mathcal M}}},\sin{\theta_{{\mathcal M}}}\sin{\phi_{{\mathcal M}}}, \cos{\theta_{{\mathcal M}}})$ and $\vec{\mu } =\mu_B( \sin{\theta_{\mu}}\cos{\phi_{\mu}},\sin{\theta_{\mu}}\sin{\phi_{\mu}}, \cos{\theta_{\mu}})$. Note that we are taking the maximum component  of the magnetic moments in complete arbitrary directions. 

We have to add now the contribution from all electrons. Given the $\phi$ dependence our elementary volume, shown in Fig.\ref{volume1} is now 

\begin{equation}
dV= b db d\phi dz
\end{equation}
and we have to integrate over $0<\phi\le2\pi$ to obtain the contribution of the full volume.

\begin{figure}[htb]
\begin{center}
\includegraphics[scale= 0.6]{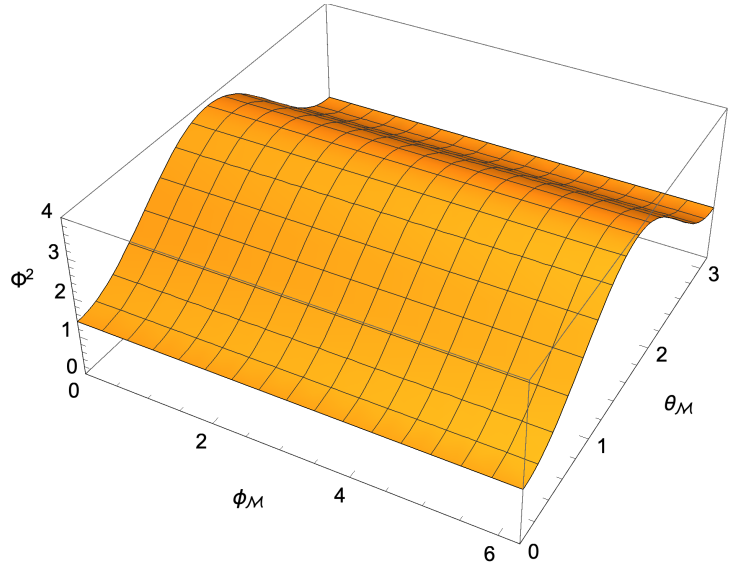} 
\end{center}
\caption{The averaged magnetic polarization function in terms of the magnetic polarization orientation of the incoming heavy particle.}
\label{PhiM2}
\end{figure}

Since the calculated energy loss will be an average value, we also average over the electron magnetic moment angles. Let us call the average magnetic polarization function in terms of the magnetic polarization of the incoming heavy particle,

\begin{equation}
\Phi_\mu^2(\theta_{{\mathcal M}},\phi_{{\mathcal M}}) =\frac{1}{2\pi^2}  \int_0^\pi d \theta_e \int_0^{2\pi} d \phi_e  \int_0^{2\pi} d\phi (\Phi_{\mu x}^2 + \Phi_{\mu y}^2),
\end{equation}
which is shown in Fig.\ref{PhiM2}.

We have to take into account the number of electrons. Recall discussion around Eqs.(\ref{eqvolume}) to \ref{energylossclassical}. Thus, our estimate for the deposited energy as a function of the incoming magnetic polarization is

\begin{equation}
\frac{dE}{dz}(\theta_{{\mathcal M}},\phi_{{\mathcal M}}) = \frac{{\mathcal M}^2 \mu_B^2 N}{2 m v^2 } \Phi_\mu^2(\theta_{{\mathcal M}},\phi_ {{\mathcal M}}) \int_{b_{min}}^{b_{max}} \frac{db}{b^5}
\end{equation}
where $N$ is the electron number density.

In order to get a feasibility study of detection of magnetic particles it is more convenient instead of calculating the deposited energy, to calculate the ratio of the deposited energy of a heavy neutral magnetic particle to that of a heavy charged particle, namely 

\begin{equation}
Ratio=\frac{\mbox{magnetic moment}}{\mbox{charge}} = \frac{{\mathcal M}^2 \mu_B^2 \Phi_\mu^2}{2 \pi Q^2 e^2} \; \frac{\frac{1}{b^4_{max}} - \frac{1}{b^4_{min}} }{\log{\frac{E_{max}}{E_{min}}}}.
\end{equation}
where $Q= Z_H e$ is the charge of the heavy particle. The impact parameter can be calculated in terms of the lab energy for one electron from Eq.(\ref{energyM})

\begin{equation}
b= \left(\frac{{8\mathcal M}^2 \mu_B^2 \left(\frac{\Phi_\mu^2}{2 \pi}\right)}{ m v^2 E}\right)^{\sfrac{1}{6}}.
\end{equation}

We average over $\phi$ to get an estimate in the equation of one electron which leads to $\frac{\Phi_\mu^2}{2 \pi}$ as written in the previous equation explicitly.  The maximum energy to leading order in $m/M$ is $E_{max} = 2 m v^2/(1-v^2/c^2)$. For $E_{min}$ it is customary to choose $10 Z$ eV~\cite{Foudas:2007cu,Ahlen:1976jw}, where $Z$ is the atomic number of the medium. In all shown results we shall use as a medium liquid Argon whose $Z=18$. Except for clarification purposes we use $\beta =v/c$ for the velocity and $m c^2$ for the electron mass.

\subsection{Energy Loss produced by the  Electric field interaction}

In this case the force $\vec{F}_E= q \vec{E}^\prime$ becomes the conventional Lorentz force

\begin{equation}
\vec{F}_E = e\beta\gamma B_y \hat{i} -e\beta \gamma B_x\hat{j} \rightarrow \frac{e}{c} \vec{v} \times \vec{B},
\end{equation}
in the low velocity limit.

Explicitly the forces become
\begin{eqnarray}
F_{E x} &=&  e \beta \gamma\left(3\frac{\vec{r}\cdot {\overrightarrow{\mathcal M} r_y}}{r^5} - \frac{{\mathcal M}_y}{r^3}\right), \\
F_{E x} &=& - e \beta \gamma \left(3\frac{\vec{r}\cdot {\overrightarrow{\mathcal M} r_x}}{r^5} - \frac{{\mathcal M}_x}{r^3}\right).
\end{eqnarray}

\begin{figure}[htb]
\begin{center}
\includegraphics[scale= 0.6]{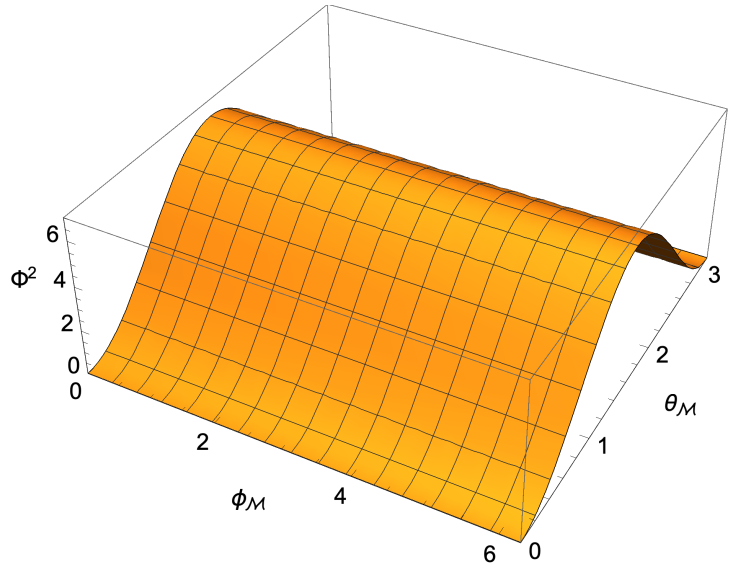} 
\end{center}
\caption{The averaged polarization function in terms of the polarization orientation of the incoming heavy particle.}
\label{PhiE2}
\end{figure}

The velocities are obtained by integrating the accelerations and lead to 

\begin{eqnarray}
v_{e x} &= &\frac{2 e  \gamma}{ m  b^2} ( 2({\mathcal M}_x \cos \phi  -  {\mathcal M}_y \sin \phi)\sin\phi  - {\mathcal M}_y)= \frac{2 e\gamma {\mathcal M}}{m b^2} \Phi_{e x},\\
v_{e y} &= &\frac{2 e  \gamma}{ m  b^2} ( 2({\mathcal M}_x \cos \phi  -  {\mathcal M}_y \sin \phi)\cos\phi  - {\mathcal M}_x)= -\frac{2 e\gamma {\mathcal M}}{m b^2} \Phi_{e y}.
\label{velocityE}
\end{eqnarray}
Therefore
\begin{equation}
E= \frac{2 e^2 \gamma^2{\mathcal M}^2 }{m b^4} (\Phi_{e x}^2 + \Phi_{e y}^2)) =\frac{2 e^2 \gamma^2 {\mathcal M}^2 }{m b^4}  \Phi_e^2.
\end{equation}
This equation defines the as before the impact parameter $b$ in terms of $E$. Integrating over $\phi$ and  $bdb$ we get

\begin{equation}
Ratio=\frac{\mbox{magnetic moment}}{\mbox{charge}} = \frac{e^2{\mathcal M}^2 \gamma^2 \beta^2 \Phi_e^2}{4 \pi Q^2 e^2} \; \frac{\frac{1}{b^2_{max}} - \frac{1}{b^2_{min}} }{\log{\frac{E_{max}}{E_{min}}}},
\end{equation}
and
\begin{equation}
b=\left(\frac{2 e^2 \gamma^2 {\mathcal M}^2 \left(\frac{\Phi_e^2}{2 \pi}\right)}{m E}\right)^{\sfrac{1}{4}}
\end{equation}
The integrated $\Phi_e(\phi_{\mathcal M},\theta_{\mathcal M})^2$ is shown in Fig. \ref{PhiE2}.

\begin{figure}[htb]
\begin{center}
\includegraphics[scale= 1.0]{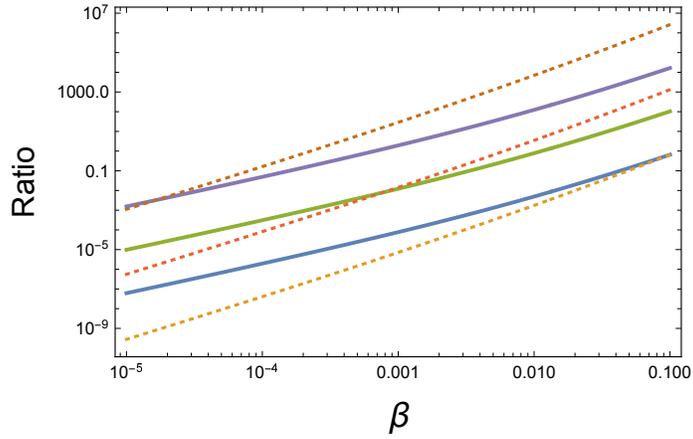} 
\end{center}
\caption{We plot the ratio of the two magnetic moment interactions  with respect to a conventional charge interaction, for the proton, as a function of velocity $\beta$ for three magnetic moments of the heavy particle. The upper curves correspond to $2000 \mu_B$, the middle curves for $\mu_B$ and the lower curves for $\mu_B/2000$, which is approximately the value of the nuclear magneton. The solid line signals the magnetic moments interaction and the dotted line the charge magnetic moment interaction.}
\label{ratiobeta}
\end{figure}

We next study the two interactions as a functions of $\beta$. We take the actual values for the electron parameters $\mu_B=0.0836$ MeV$^{-1}$, $m=0.511$MeV, $ q = - e = -\sqrt{1/137}$ and plot the ratio for three magnetic moments of the heavy particles. We average over $\Phi_{\mathcal M}$ and $\theta_{\mathcal M}$ getting for the average values of ${\Phi_\mu}^2=2.618$ and of ${\Phi_e}^2 = \pi$. The result is shown in Fig. \ref{ratiobeta}. We see that for the higher magnetic moments the charge magnetic moment interaction dominates due to its softer radial dependence for almost all the range of $\beta$ studied. For the lower magnetic moments, at the level of nuclear magnetons, the magnetic moments interaction dominates due to the velocity dependence. The $Z_H$ dependence is easy to understand it simply divides the Ratio by $Z_H^2$.

\begin{figure}[htb]
\begin{center}
\includegraphics[scale= 1.0]{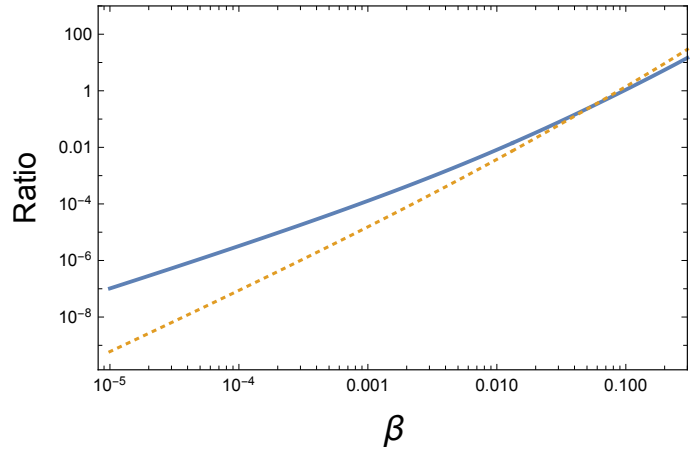} 
\end{center}
\caption{We plot the ratio of the two magnetic moment interactions for the neutron with respect to a conventional charge interaction, for the alpha particle, as a function of velocity $\beta$. The solid line corresponds to the magnetic moments interaction and the dotted line to the charge magnetic moment interaction.}
\label{rationeutron}
\end{figure}

As an application of our calculation to a well known systems we plot the comparison between the ratio of the energy loss of  a neutron to that of an alpha particle in Fig.\ref{rationeutron}. In this case the magnetic moment interaction dominates over the whole range of velocities studied due to the smallness of the nuclear magneton .

\subsection{Energy Loss associated with the full interaction}

The aim of this work is to analyze the energy loss of monopolium and for that purpose we are going to implement the full interaction. The procedure consists in adding all forces acting of the heavy particle and constructing from them the velocity which leads immediately to the lost energy.

The total velocity requires the addition of Eqs.(\ref{velocityM}) and (\ref{velocityE}) and can be written as 

\begin{eqnarray}
v_{e x} & = &\frac{4{\mathcal M}\mu_b}{\beta b^3m} \Phi^\mu_x +\frac{2 e \gamma {\mathcal M}}{m b^2} \Phi^e_x\\
v_{e x} & = -&\frac{4{\mathcal M}\mu_b}{\beta b^3m} \Phi^\mu_y +\frac{2 e \gamma {\mathcal M}}{m b^2} \Phi^e_y
\end{eqnarray}
where we use the natural system of units and $\Phi^\mu_i$ and $\Phi^e_i$ are the polarization functions appearing in Eqs.(\ref{velocityM}) and (\ref{velocityE}) when substituting  $\overrightarrow{{\mathcal M}} = {\mathcal M}( \sin{\theta_{{\mathcal M}}}\cos{\phi_{{\mathcal M}}},\sin{\theta_{{\mathcal M}}}\sin{\phi_{{\mathcal M}}}, \cos{\theta_{{\mathcal M}}})$, $\vec{\mu } =\mu_B( \sin{\theta_{\mu}}\cos{\phi_{\mu}},\sin{\theta_{\mu}}\sin{\phi_{\mu}}, \cos{\theta_{\mu}})$.

The energy transmitted to the electron for fixed $\theta_\mu, \phi_\mu,\theta_{\mathcal M},\phi_{\mathcal M}, \phi$ and $b$ becomes

\begin{equation}
\frac{2 {\mathcal M}^2}{m b^4} \sum_{i =x,y} \left(\frac{2 \mu_B}{\beta b} \Phi^\mu_i + e \gamma \Phi^e_i)^2\right).
\label{energy}
\end{equation}

We next integrate this expression $\frac{1}{2 \pi^2} \int d \theta_\mu d \phi_\mu d \phi$. The cross terms  $\frac{1}{2\pi^2} \int \Phi^\mu_i\Phi^e_i d \theta_\mu d \phi_\mu d \phi$ vanish, the energy transferred to all electrons with fixed $\theta_{\mathcal M},\phi_{\mathcal M}$ and $b$ is

\begin{equation}
E= \frac{8 {\mathcal M}^2\mu_B^2}{m\beta^2}\left(\frac{\frac{\Phi_\mu^2}{2 \pi}}{b^6} + \frac{e^2 \beta^2\frac{\Phi_e^2}{2 \pi}}{4\mu_B^2}{b^4}\right),
\end{equation}
where $\Phi^2 \sim \frac{1}{2\pi^2}\int \sum_i \Phi_i^2 d \phi_\mu d \theta_\mu d \phi$, with $i=x,y$. Figs. \ref{PhiM2} and \ref{PhiE2} show the structure of the polarization functions after this integration.

This equation relates $E$ to $b$ and will be used to eliminate $b$ in terms of $E$ in the final calculation. However for one electron we average the polarization function over $\phi$, thus the contribution will be in that case $\frac{\Phi^2}{2 \pi}$.
Following the procedure outlined before we have next to perform the integration over $\int b d b$ to obtain the energy for all the electrons in the $b, b+ db$ cylinder. Thus the energy lost by the particle as it travels an infinitesimal $dz$ is

\begin{equation}
\frac{d E}{dz} =  \frac{2 {\mathcal M}^2\mu_B^2}{m\beta^2}\left( \Phi_\mu^2 \left(\frac{1}{b_{max}^4}-\frac{1}{b_{min}^4} \right)+ \frac{e^2 \beta^2\Phi_e^2}{2\mu_B^2}\left(\frac{1}{b_{max}^2}-\frac{1}{b_{min}^2} \right)\right).
\end{equation}

The next step is to use Eq.(\ref{energy}) to relate $b$ to the energy. In order to get estimates we shall average over polarization angles and $\phi$. The result becomes

\begin{equation}
\sigma^3 -P\sigma -Q = 0
\end{equation}
where  
\begin{eqnarray}
b&=&\frac{\sqrt{\sigma}}{m}, \nonumber\\
P&=&\frac {2{\mathcal M}^2 e^2 \frac{\Phi_e^2}{2\pi}m^3}{E}, \nonumber\\
 Q&=& \frac{8{\mathcal M}^2 \mu_B^2 \frac{\Phi_\mu^2}{2\pi} m^5}{\beta^2 E}.
 \end{eqnarray}
The coefficients and the variables in the Cardano polynomial have been multiplied by appropriate powers of $m$ to have all of them adimensional. This equation defines $b_{max},b_{min}$ as a function of $E_{max},E_{min}$.

\begin{equation}
Ratio= \frac{2 {\mathcal M}^2\mu_B^2 \left( \Phi_\mu^2 \left(\frac{1}{b_{max}^4}-\frac{1}{b_{min}^4} \right)+ \frac{e^2 \beta^2\Phi_e^2}{2\mu_B^2}\left(\frac{1}{b_{max}^2}-\frac{1}{b_{min}^2} \right)\right)}{4 \pi Q^2 e^2 \log{\frac{E_{max}}{E_{min}}}}
\label{fullratio}
\end{equation}

\begin{figure}[htb]
\begin{center}
\includegraphics[scale= 1.0]{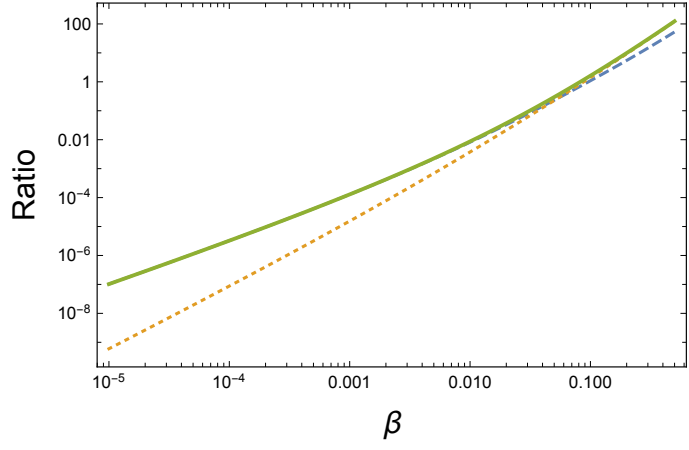} 
\end{center}
\caption{We plot the ratio of the full interaction for the neutron with respect to a conventional charge interaction for the alpha particle, as a function of velocity $\beta$. The dashed line shows the result of only the magnetic moments interaction and the dotted line that of the charge magnetic moment interaction.}
\label{rationeutronfull}
\end{figure}

As an application of the procedure we show in Fig. \ref{rationeutronfull} the full calculation (thick)  for the neutron and we compare it with the results of the magnetic moment interaction (dashed) and electric charge magnetic moment interaction (dotted). The magnetic moments interaction is barely noticed, because it is dominant for the neutron magnetic moment and the range of velocities studied, and therefore almost coincides with the full calculation as can be confirmed by comparing this figure with Fig. \ref{rationeutron}. We have extended in this case $\beta$ to $0.3$ simply to make manifest the difference between the full and the magnetic moments interaction.

Having finished the theoretical developments leading to the  full interaction ratio we proceed to apply our result to monopolium.

\section{Analysis for Monopolium}
\label{monopolium}

In this section we apply the previous developments to the excited  high lying Rydberg maximally deformed states of monopolium. In the conventional studies of energy loss the value of $E_{min}$ used  is $ \sim 10 Z$ eV, where  $Z$ is the atomic number of the atom in the medium and as mentioned before we use liquid Argon, $Z=19$.

\begin{figure}[htb]
\begin{center}
\includegraphics[scale= 0.9]{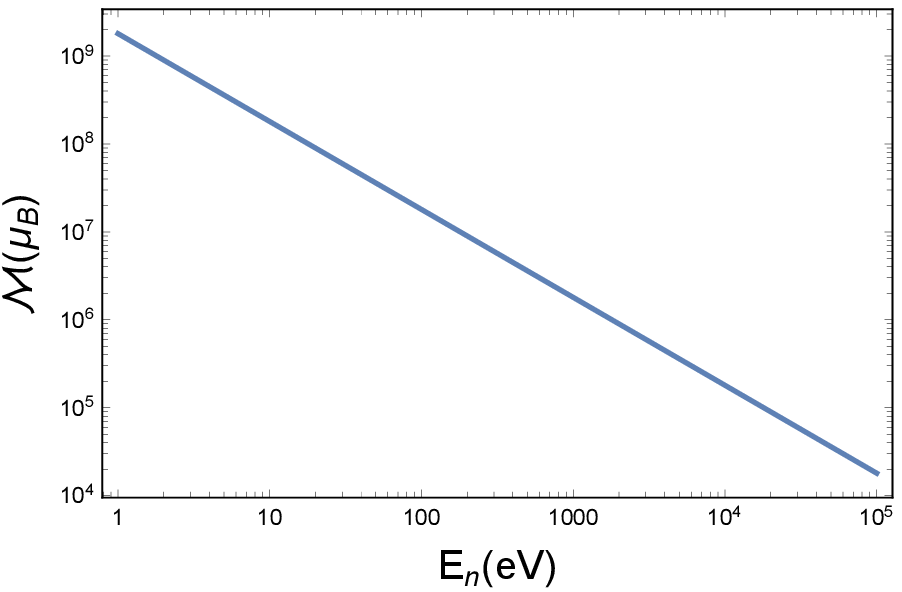} \hspace{0.5cm} \includegraphics[scale= 0.9]{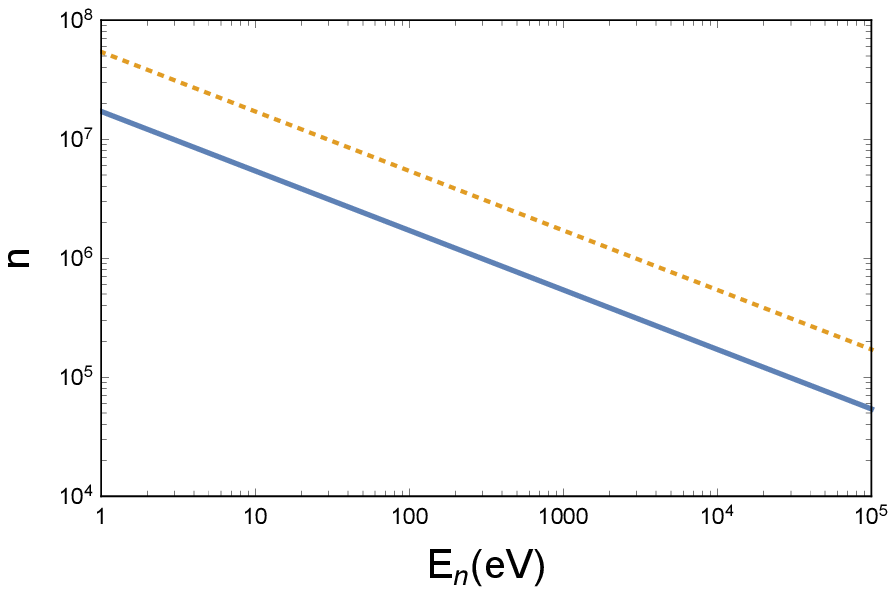} 
\end{center}
\caption{Left:We plot the relation between the corresponding magnetic moment in Bohr magnetons and their corresponding absolute value of the monopolium's $n$th state binding energy. Right: we show the relation between the binding energy and the principal quantum number of the state for two monopole masses $M=1$ TeV (solid) and $M=10$ TeV (dotted).}
\label{relation}
\end{figure}

In monopolium  the high $n$ states are those of a magnetic Coulomb potential  $\frac{g}{r}$ and therefore for each $n$ there is a degeneracy in $l$ in the absence of external fields. Consequently even with a weak magnetic or electric field, which will always be present in our experiments,  there will be a complete $l$ mixing. Calculating the dual Stark effect in parabolic coordinates, and comparing with the energy shift of a magnetic dipole field, we obtain the magnetic moment of a Rydberg state in the presence of a magnetic field. For  high $n$ the maximal magnetic moment is \cite{Damburg_1979,Gallagher_1988,Hogan:2013qla,Alonso:2018sun,Fanchiotti:2022xvx},
 
\begin{equation} 
{\mathcal M_n} \sim   \frac{3}{2} n^2 g\; r_{Bohr} \sim g < n 0| r |0 n> \sim \frac{3}{32\,\alpha^{\sfrac{3}{2}}\,E_n},
\label{magneticmoment}
\end{equation}
where $E_n$ is the absolute value of the $n$th state binding energy.   These expressions correspond to the large $n$ limit. 

\begin{figure}[htb]
\begin{center}
\includegraphics[scale= 0.9]{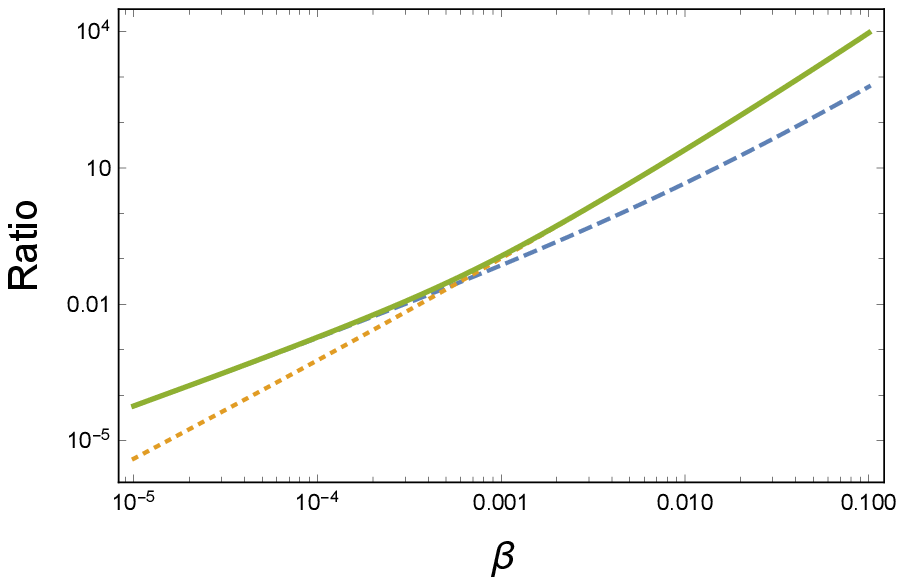} \hspace{0.5cm} \includegraphics[scale= 0.9]{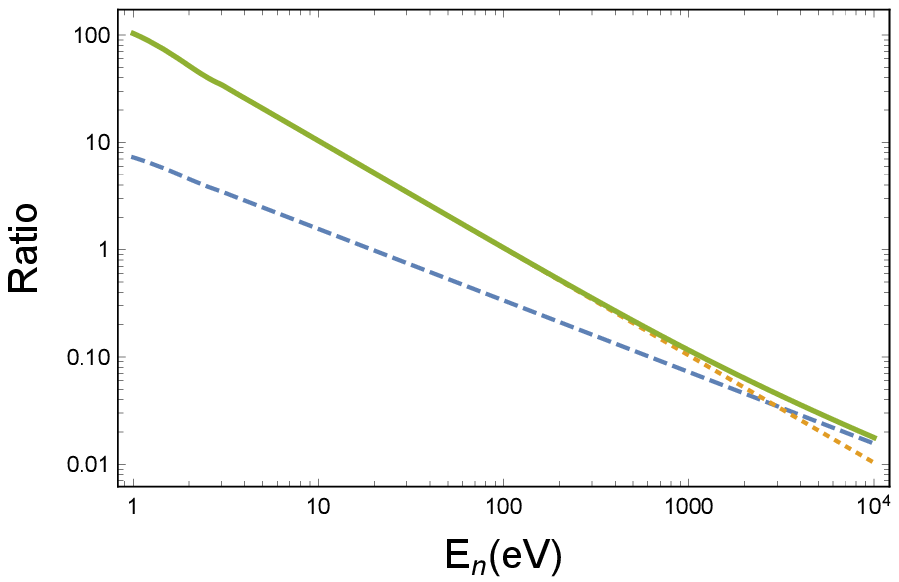} 
\end{center}
\caption{Left: Ratio of energy losses between monopolium and the proton for $E_n =1000$ eV as a function of $\beta$. Right: Ratio of energy loses between monopolium and the proton for $\beta=0.001$ as a function of binding energy $E_n$ in eV. The solid curve represents the full calculation, the dashed curve the ratio taking into account only the magnetic moment interaction, and the dotted curve the ratio taking into account only the charge magnetic moment interaction.}
\label{ratio}
\end{figure}

We show in Fig. \ref{ratio} the largest possible ratio of the energy loss of  monopolium by comparing the monopolium energy loss with that of a proton $Z_H=1$. The ratio for any other ion is obtained dividing by its $Z_H^2$. The remaining inputs are Bohr's magneton given by $\mu_B \sim 0.0837$MeV$^{-1}$, the electron mass $m\sim0.511$ MeV, and the unit charge $e\sim\sqrt{1/137}=0.0854$

We express our results as a function of the absolute value of the binding energy $E_n$. For physical purposes we show in  Fig. \ref{relation} the relation between  $E_n$ in eV and the  maximal magnetic moment  ${\mathcal M}$ of that $n$th level. In the right plot of the same figure we show the correspondence between the binding energy and the principal quantum number for monopole masses of $1$ TeV (dashed) and $10$ TeV (dotted) as given by Eq. (\ref{En}),

\begin{equation}
E_n= \frac{3 M}{48 \alpha^2 n^2},
\label{En}
\end{equation}
where $M$ is the mass of the monopole.

In Fig. \ref{ratio} we show (left) the ratio of energy losses between a high lying monopolium Rydberg state of $E_n = 1000$ eV and a proton The right figure shows the same ratio of energy losses as a function of  binding energy for $\beta=0.001$. For larger $\beta$s the charge magnetic moment interaction will dominate and for smaller the magnetic moments interaction will be dominant. These results point out that if highly deformed monopolia would reach a liquid $Ar$ detector they would produce a  ionization of the same order of magnitude as that of light-medium ions. 

\begin{figure}[htb]
\begin{center}
\includegraphics[scale= 0.95]{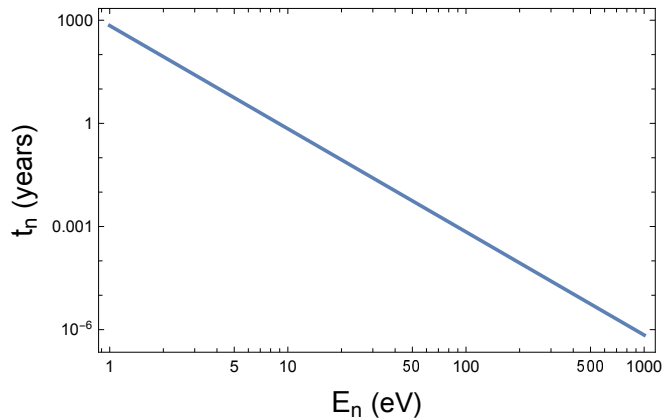} 
\end{center}
\caption{Lifetime of monopolium Rydberg states in years as a function of binding energies.}
\label{life}
\end{figure}

One might be surprised by the extremely large magnetic moments required for the observation corresponding to huge principal quantum numbers. Will these very high $n$ monopolium Rydberg states reach detectors on Earth? The  Rydberg states with long lifetimes and large magnetic moments are lightly bound monopole-antimonopole pairs with binding energies of the order of eVs.  The small Parker bound, an upper bound on the density of magnetic monopoles~\cite{Turner:1982ag}, can be obviated because we are dealing with bound states~\cite{Dicus:1983ry}. The Parker bound is based on the monopoles taking energy from the intergalactic magnetic field. However, monopolia below the dissociation energy do not take energy from the magnetic field since magnetic moment interactions at large distances are much weaker than magnetic charge interactions.
Moreover, the highly deformed Rydberg states which are in high $l$ states, therefore with vanishing wave function at the origin, have the monopole and antimonopole far away from each other and therefore do not annihilate easily as it is the case of the low lying states. They will de-excite by spontaneous emission of photons to the lower levels where they will annihilate either as S states or to non leading order  as $l\ne 0$ state decays. In order to get an estimate for spontaneous emission as a function of $n$ using the naive Bohr model one gets~\cite{Dicus:1983ry},

\begin{equation}
\Gamma^{se}_n(E) = 4096 \,\beta_n^4 \, \alpha \,m \left(\frac{E_n}{m} \right)^3.
\label{spontaneousemission}
\end{equation}
This width's $n$ dependence dependence goes like $1/n^6$, therefore lifetimes of these Rydberg states increase notably with $n$ as seen in Fig. \ref{life}.  By astronomical standards the distances they can travel are small and therefore they will have to be produced by high energy photons in our vecinity.

Similar calculations can be extended easily to Kaluza-Klein (KK) monopolium as was shown in ref. \cite{Vento:2007vy}, but in this case since the monopoles do not decay classically their lifetime is enormous and therefore they could reach us from all parts of the universe, most probably quite cold and therefore our non relativistic result applies to them very nicely. To translate to that calculation we have simply to transfer the results to magnetic moments and apply the formula for magnetic moments to the magnetic moment of that of the KK monopolia in terms of the distance between the poles \cite{Gross:1983hb,Vento:2007vy}.

\section{Concluding remarks}

Monopolium, a bound state of monopole-antimonopole, has no magnetic charge and in its ground state no magnetic moment, being therefore very difficult to detect directly. In conventional gauge theories it has the quantum numbers of the vacuum and annihilates into photons, and these disintegrations have been intensively studied \cite{Epele:2007ic,Epele:2008un,Epele:2012jn,Barrie:2016wxf,Fanchiotti:2017nkk,Barrie:2021rqa}. In Kaluza-Klein theories monopolium is classically stable and therefore very long lived \cite{Gross:1983hb,Vento:2020vsq} but its detection problems still persist since it only interacts gravitationally. What happens with excited monopolium states? This paper deals with the study of the excited states of monopolium. Excited states have in the presence of even weak magnetic fields magnetic moments and their interaction with magnetic and electric fields is known.  Recently we used a theoretical analysis of magnetic dipole detection to learn about excited monopolium states \cite{Fanchiotti:2022xvx}. In this work we have analyzed the capacity of monopolium to produce ionization when traversing a medium by calculating its energy loss when colliding with  atomic electrons. The idea behind this investigation is to find out if TPCs \cite{Nygren:2018sjx} could be used for detecting monopolium.  We have calculated the energy loss of monopolium in an $Ar$ medium by a classical analog of the Bethe-Bloch formula. The analogous calculation for a charge particle and for a monopole agrees well with the Bethe-Bloch equation \cite{Bethe:1930ku,Bloch:1933zza,Ahlen:1976jw} in the non relativistic limit, and thus we expect that the same will happen for monopolium.  We have seen that in those kinematical regions where the density dependence is not important and for non relativistic particles the ionization of monopolium excited Rydberg states is of the same magnitude as that to medium-light ions, the result being  independent of the heavy particle mass (recall Fig. \ref{ratio}). The Bethe-Bloch formulation is an average description and therefore our result does not determine the precise ionization of one monopolium event but certainly determines the order of magnitude of the effect from the perspective of the sensitivity of the detectors. 

\section*{Acknolwedgements}
HF and CAGC were partially supported by ANPCyT, Argentina. VV was
supported by MCIN/AEI/10.13039/501100011033, European Regional Development Fund Grant No. PID2019-105439 GB-C21 and by GVA PROMETEO/2021/083. VV acknowledges correspondence with Alan Chodos and David Nygren which motivated this investigation, and useful discussions on monopole ionization with Laura Patrizii.

\end{document}